\documentclass{aa}  

\usepackage{graphicx}
\usepackage{txfonts}
\usepackage{lipsum}
\usepackage{subcaption}         % necessary for continued figures, example in section 3
\usepackage{lscape}             % to rotate a single page table, example in appendix.
\usepackage{placeins}           % useful with \FloatBarrier, to keep 
\begin{document}

%%%%%%%%%%%%%%%%%%%%%%%%%%%%%%%%%%%%%%%%
% if you use custom commands in your title,
% ensure to check your title when submitting!
%%%%%%%%%%%%%%%%%%%%%%%%%%%%%%%%%%%%%%%%
\title{ON FLYING NEAR THE BASE OF A PSEUDO-STREAMER}

\subtitle{}

%%%%%%%%%%%%%%%%%%%%%%%%%%%%%%%%%%%%%%%%
% Please separate each author with the \and command
%
% Please do not include ORCIDs next to author names.
% Only ORCIDs authenticated by individual authors in EDPS
% editorial system will be taken into account.
% ORCIDs included here will be removed.
%%%%%%%%%%%%%%%%%%%%%%%%%%%%%%%%%%%%%%%%

   \author{F.S. Mozer\inst{1}\fnmsep\thanks{Corresponding author: forrest.mozer@gmail.com}
        \and O. V.  Agapitov\inst{1}
        \and K.-E. Choi\inst{1}
        \and A. Voshchepynets\inst{2}
        }

   \institute{Space Sciences Laboratory, University of California, Berkeley, CA. USA
   \and Dept. of System Analysis and Optimization Theory, Uzhhorod National University, Uzhhorod, Ukraine}

   \date{}

% \abstract{}{}{}{}{}
% 5 {} token are mandatory
 
  \abstract
   {Near the 10 solar radius perihelion of Parker Solar Probe orbit 24, a confined region was encountered containing an enhanced plasma density (25,000 $\text{cm}^{-3}$) and broadband electrostatic waves. The solar wind velocity (200 km/sec) and ion temperature (25 eV) were significantly reduced compared to their values in the ambient solar wind.  Because the polarity of the radial magnetic field did not change sign on the two sides of the crossing and the crossed region contained a double-peaked plasma structure, the spacecraft must have passed near the base of a pseudo-streamer.  In the plasma frame, an electric field as large as 400 mV/m was detected during the crossing.  This field was not associated with an $E\times B$ drift, because it was observed in the plasma rest frame where such a drift had been removed.  Instead, this field was balanced by Generalized Ohm's Law terms  that contain a $J\times B$ force, pressure gradients, and turbulence, all of which were observed to be present.  If the $J \times B$ term was dominant,  a current of about  $1 \text{ mA/m}^2$ was present.   Turbulent plasma density fluctuations ($\Delta n /n$), as large as 0.3, suggest that the resistive term in the Generalized Ohm's Law (GOL) was significant.  In addition, both the density and the electric field had non-zero slopes as functions of time, suggesting that the pressure gradient term in the GOL also mattered.  This large electric field in the plasma rest frame may be the first and largest such field ever reported.}

\keywords{pseudo-streamer physics}
\maketitle   
\nolinenumbers

%%%%%%%%%%%%%%%%%%%%%%%%%%%%%%%%%%%%%%%%%%%%%%%%%%%%%%%%%%%%%%
\section{Introduction}
Coronal streamers are frequently detected in remote observations of the solar corona.  They are confined spatial structures containing enhanced plasma densities, slow solar wind, and cold perpendicular ions \citep{Koutchmy2005,Morgan2020}. Coronal streamers fall into two primary topological categories: helmet streamers (or bipolar streamers) and pseudo-streamers (or unipolar streamers). The distinction is defined by the magnetic polarity of the adjacent coronal regions. Helmet streamers separate coronal regions of opposite polarity and are characterized by a large sheet of closed, trans-equatorial magnetic field lines that ultimately form the Heliospheric Current Sheet (HCS) beyond the Sun \citep{wilcox1967,WangSheeleyJrRich2007,Owens2013}. 

In contrast, pseudo-streamers separate coronal regions of the same magnetic polarity and do not feature a heliospheric current sheet \citep{Wang2019,Abbo2016}. This fundamental difference in magnetic structure leads to distinct plasma properties and solar wind characteristics. Pseudo-streamer plasma sheets are thought to exhibit lower internal electron densities and are generally described as more quiescent than helmet streamers, showing a steady outflow without the characteristic density spikes or plasma blobs associated with helmet streamer activity \citep{WangSheeleyJrRich2007,Wang2012,Abbo2015}.  Furthermore, in the literature, helmet streamers are unambiguously linked to the slowest component of the solar wind, while pseudo-streamers are typically associated with wind speeds in the intermediate range, spanning from approximately 320 to 600 km/s \citep{Wang2012,Wang2019,Wallace2020}. The precise contribution of these structures to the acceleration of the solar wind is crucial, as they are known to possess multiple topological elements that enable energy release through interchange reconnection  \citep{PellegrinFrachon2023, Owens2013}.  For the event discussed in this paper, the relationship between the pseudo-streamer and the large-scale solar wind structure has been described \citep{Bale2025}.

From the solar surface outward, the parts of a pseudo-streamer consist of a base, a stalk, a topological boundary, and an open field line region. The base is a closed magnetic field line region thought to extend over several solar radii and is the foundation of the structure, consisting of magnetic field lines that loop back down to the Sun's surface. These closed loops hold the dense, hot plasma associated with bright features visible in coronal images. Despite their recognized importance as persistent structures, direct, in situ measurements deep within the plasma sheet of a pseudo-streamer have previously been inaccessible. The Parker Solar Probe (PSP), has now reached the inner heliosphere, offering an unprecedented opportunity to fly near these magnetically defined boundaries. Understanding the physical parameters, including the full velocity profile, turbulence characteristics, thermal state, fields, and currents, is essential to validating magnetohydrodynamics (MHD) models and resolving the origin of the slow and intermediate solar wind.  In this paper, PSP observations near the base of a pseudo-streamer are presented.

\section{Data}
The electric and magnetic field data came from the FIELDS instruments \citep{Bale2016} while the plasma data was produced by the SPAN instruments \citep{Whittlesey2020}. All data are presented in the PSP spacecraft frame whose positive Z-direction points to the Sun. 

% --- Figure 1 ---
\begin{figure}[t!]
    \centering
    \includegraphics[width=\linewidth]{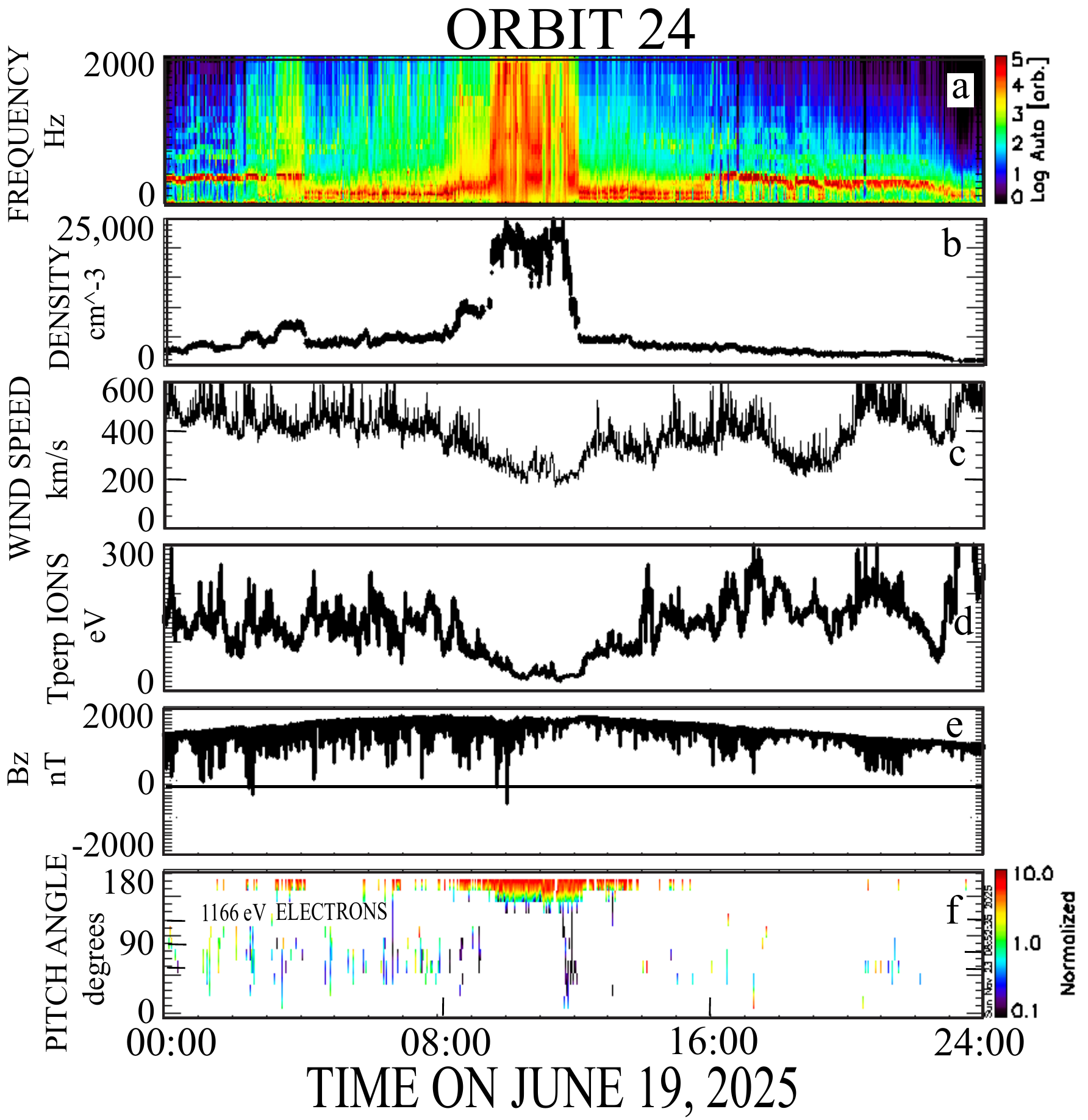}
    \caption{(1a) the spectrum of the electric field near the PSP perihelion of 10 solar radii; (1b) the plasma density, which had a major enhancement at the time and location that the broadband waves appeared in (1a); (1c) a slow solar wind ; (1d) the ion perpendicular temperature was low at this time; (1e) the radial component of the magnetic field had the same orientation on both sides and through the event except for a short time interval during the event; (1f) the electron pitch angle distribution at an energy of 1166 eV, which shows that an outflowing beam of keV electrons appeared through the event.}
    \label{fig:fig1}
\end{figure}
Figure 1 displays PSP measurements near 10 solar radii on encounter 24.  During a $\sim$three-hour interval near perihelion, in Figure 1a there were major changes in the emission spectral characteristics from narrow-band waves to wide-band emissions. Over most of the region, there were few hundred Hz, narrow-band, triggered ion acoustic waves that have been described. \citep{Mozer2021, Mozer2022_Density, Mozer2023, Mozer2025a}.  During the shorter time interval of the broad band waves, as seen in Figure 1b, the plasma density was almost an order of magnitude increased relative to ambient values (from $\sim3-5 \times 10^3$ to $2.0-2.5 \times 10^4 \text{ cm}^{-3}$) while the solar wind flow was slow ($200 \text{ km/sec}$) (Figure 1c) and the perpendicular ion temperature was a factor of more than two lower than in the typical solar wind (Figure 1d). The radial component of the magnetic field, illustrated in Figure 1e, was positive on both sides of the region and had a brief negative excursion during the large density interval.  The 1166 eV electron pitch angle distribution (the high energy component of the electron suprathermal population) in Figure 1f had a beam-like distribution near 180\textsuperscript{o}, which means that the electron beam direction was opposite to the direction of the magnetic field or away from the Sun. 

% --- Figure 2 ---
\begin{figure}[t!]
    \centering
    \includegraphics[width=.9\linewidth]{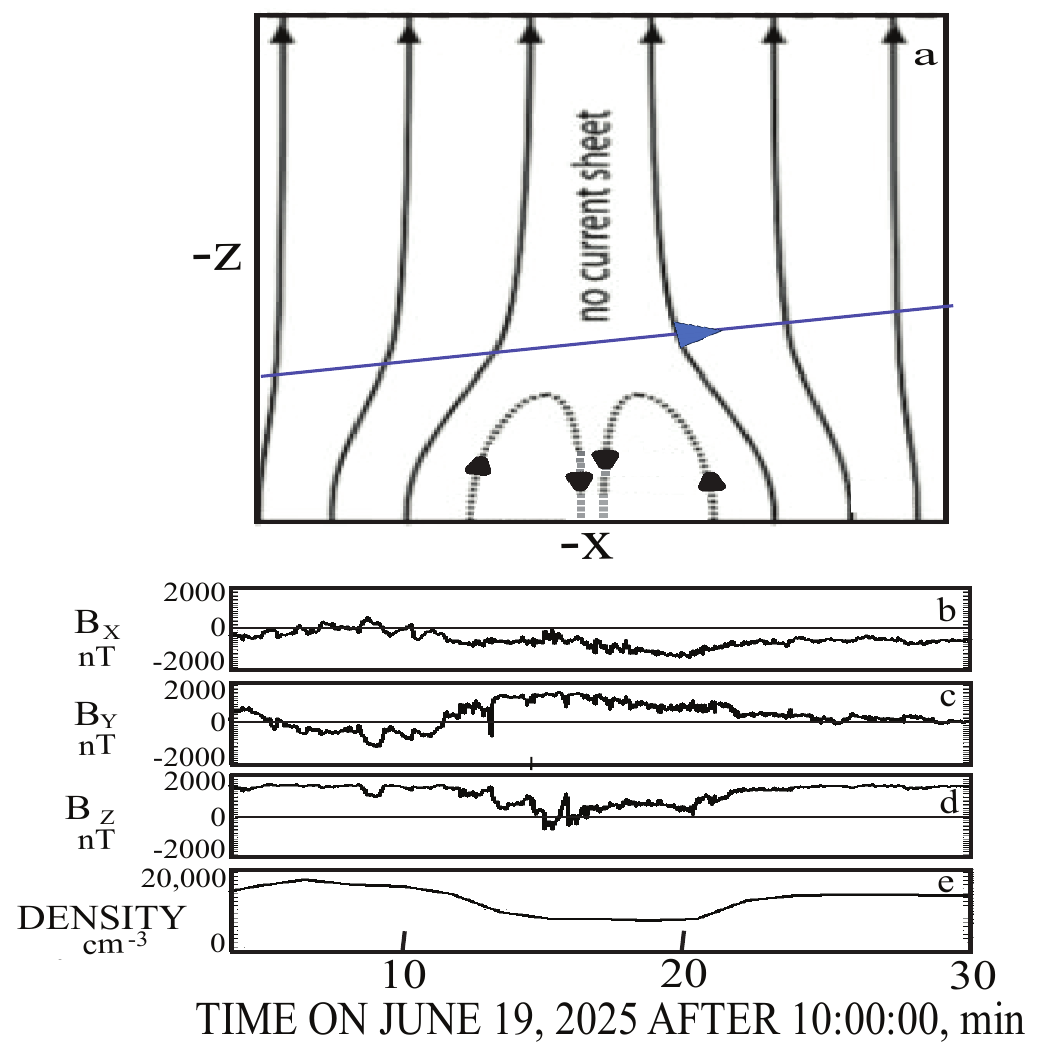}
    \caption{(2a) The magnetic field geometry near a pseudo-streamer, a modified Figure taken from \citep{Crooker2014}. It illustrates the double structure at the base of the pseudo-streamer and shows (in purple) a plausible spacecraft trajectory near it;  (2b), (2c), and (2d) measured magnetic field components; (2e) the plasma density}
    \label{fig:fig2}
\end{figure}
The plasma density shown in Figure 1b had a two-peak distribution. This is consistent with and required for a pseudo-streamer, as illustrated by the model of its magnetic field in Figure 2a. This model also requires a reversal of the radial component of the magnetic field between the two peaks when the trajectory is near the base of the pseudo-streamer. Examination of Figure 1e shows that the data contained a narrow, negative (radial) $B_Z$ component. This region is expanded in Figures 2b through 2d and a plausible trajectory of the spacecraft through this region is given as the purple curve in Figure 2a. On this trajectory, the radial-component of the magnetic field, initially positive, tends to go through zero to become negative, and then returns to being positive through the rest of the trajectory. In addition, $B_Y$ becomes initially negative and then positive in the model trajectory. These are the $B_Z$ and $B_Y$ behaviors illustrated in Figures 2d and 2c. Meanwhile, the plasma density of Figure 2e was minimum in the vicinity of the field reversal, as expected from the trapping of the plasma in the double peak distribution. This agreement of the data with the two-peak geometry is a remarkable confirmation that the spacecraft passed near the base of a pseudo-streamer.

% --- Figure 3 ---
\begin{figure}[t!]
    \includegraphics[width=\linewidth]{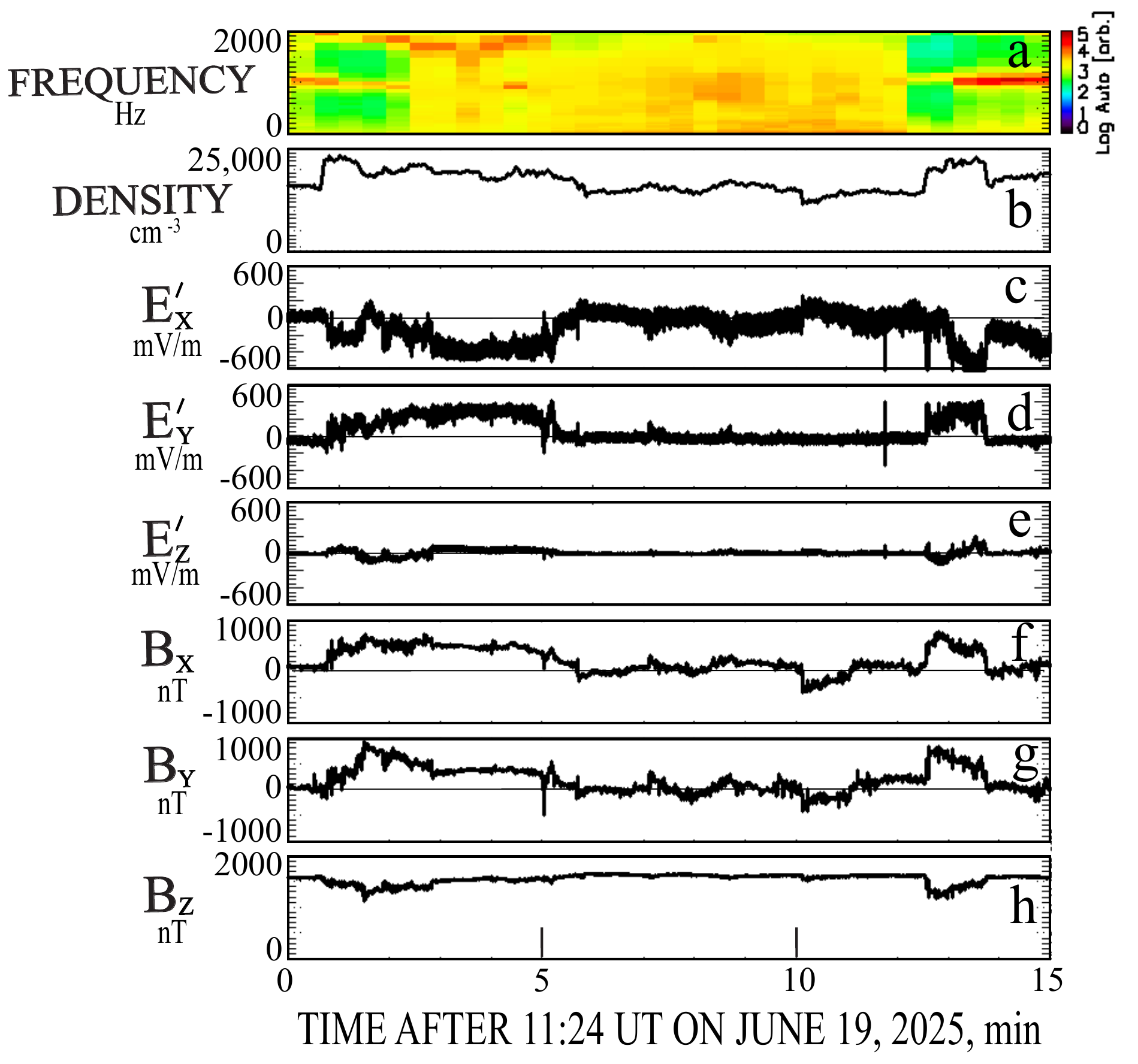}
    \caption{The electric and magnetic fields measured in the plasma frame of rest inside the pseudo-streamer: (3a) gives the electric field wave spectrum, (3b) gives the plasma density; (3c), (3d), and (3e) give the prime electric field; (3f), (3g), and (3h) give the magnetic field.}
    \label{fig:fig3}
\end{figure}

 Given that the spacecraft was near the base of a pseudo-streamer, the electrodynamics of the plasma was investigated through examination of the fields. The measured $E_X$ and $E_Y$ were transformed into the plasma rest frame by removing vxB, where v is the solar wind velocity.   In this prime frame, there were  electric fields as large as 400 mV/m, (see Figures 3c and 3d) that occurred during maxima of the magnetic fields shown in Figures 3f, 3g, and 3h, as well as maxima in the plasma density of Figure 3a.  These large electric fields were not associated with an ExB drift because those parts of the electric field were removed when the data was transformed into the plasma rest frame.  Instead, these large electric fields must have been force-balanced by terms in the Generalized Ohm's Law (GOL) which is
\begin{equation}
\mathbf{E}' = \mathbf{E} + \mathbf{v} \times \mathbf{B} = \eta \mathbf{J} + \frac{\mathbf{J} \times \mathbf{B}}{ne} - \frac{\nabla \cdot \mathbf{P}_e}{ne} + \frac{m_e}{ne^2} \frac{\partial \mathbf{J}}{\partial t} ~,
\end{equation}
where the $\mathbf{v} \times \mathbf{B}$ term is less than 20\% of $\mathbf{E}$ and $\mathbf{v}$ 
is the plasma bulk velocity, $\eta$ is the collision resistivity, $\mathbf{J}$ is the
current density, $n$ is the plasma density, and $\nabla \cdot \mathbf{P}$ is the pressure gradient.  The last term in the GOL equation is neglected in the following discussion because it was negligible.

Additionally, the unmeasured $E_Z$ was reconstructed by assuming that the parallel electric field was zero or that
\begin{equation}
E_Z = \frac{-E_X B_X - E_Y B_Y}{B_Z}
\end{equation}
Its value is illustrated in Figure 3e.

If  the $\mathbf{J} \times \mathbf{B}$ term were the dominant term on the right side of the GOL, the perpendicular current density near the pseudo-streamer can be estimated after making three  coordinate rotations at every data point; first rotating about the X-axis to make $B_Y = 0$, then rotating about the Y-axis to make $B_X=0$, and lastly rotating about the Z-axis to make $E_Y=0$.  In this frame, the non-zero field components are $E_X$ and $B_Z$, illustrated in Figures 4c and 4d.  Combining these quantities with the plasma density of Figure 4a allows the calculation of the perpendicular current density of Figure 4b to be approximately $1 \text{ mA/m}^2$.   (The parallel current density was also non-zero, as may be seen from the electron pitch angle distribution of Figure 1f.)  This is an overestimate of the actual perpendicular current density because it neglects the  GOL collisional resistivity and pressure gradient terms. That the resistivity term was also non-zero is shown by the large amplitude density fluctuations $\Delta n/n$, reaching as high as $0.3$,  and illustrated in Figure 5b.  This data was obtained by estimating the high frequency plasma density of Figure 5a by fitting the  60-second-averaged spacecraft potential to the logarithm of the similarly averaged plasma density and using the fitting coefficients to expand the high rate spacecraft potential measurements into a proxy for the plasma density and its fluctuations.

That the pressure gradient term may have been non-zero is indicated by the finite slope  with time of the density plot in Figure 3b during the two time intervals (one near the beginning and one near the end of Figures 3c and 3d) when the electric field was non-zero.
% --- Figure 4 ---
\begin{figure}[t!]
    \includegraphics[width=\linewidth]{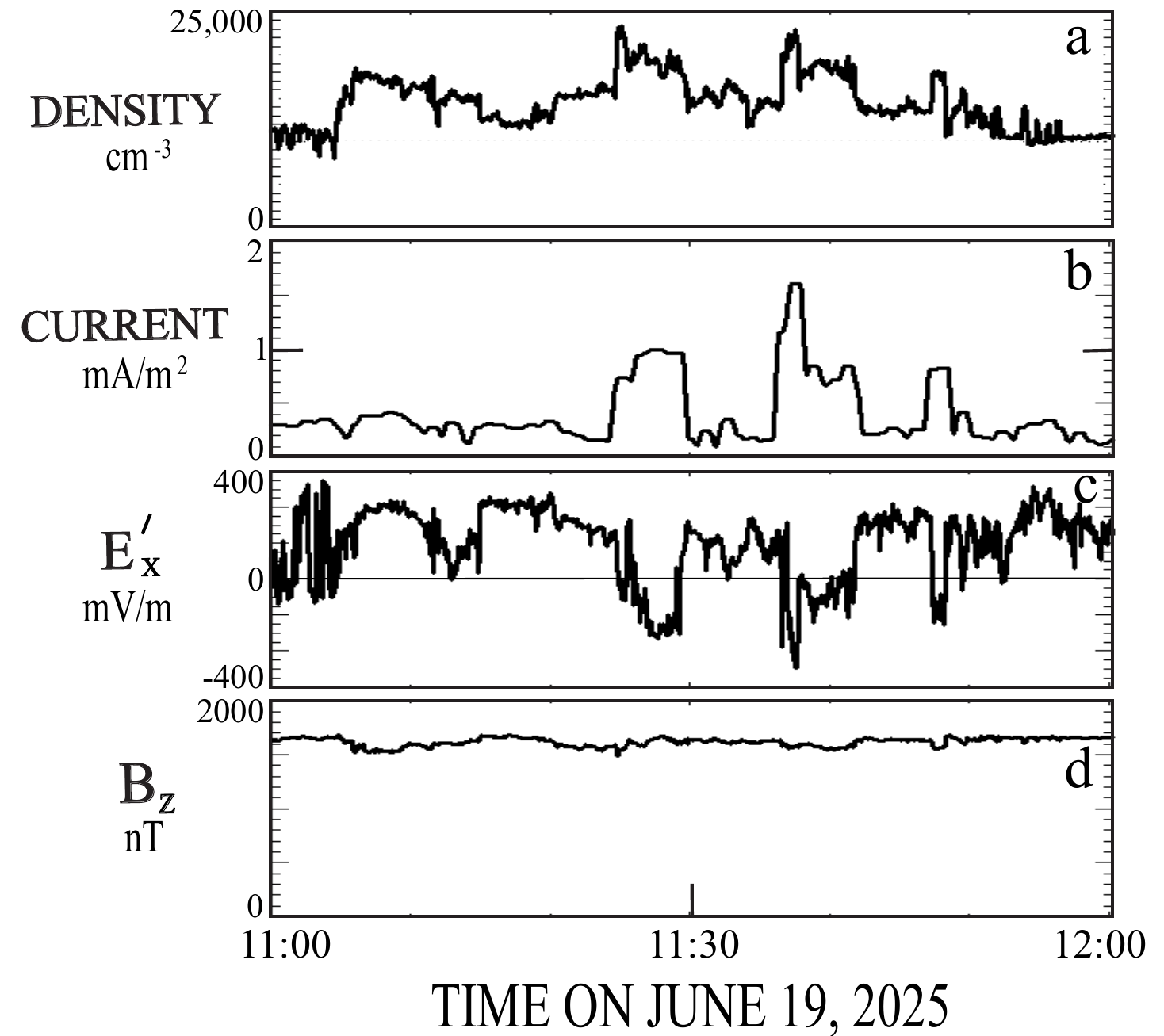}
    \caption{(4a) The plasma density; (4c)  the total electric field in the prime frame: (4d) the total magnetic field. (4b)  the current density obtained from assuming JxB/ne=E and using the data in (4a), (4c), and (4d). }
    \label{fig:fi4}
\end{figure}
% --- Figure 5 ---
\begin{figure}[t!]
    \centering
    \includegraphics[width=\linewidth]{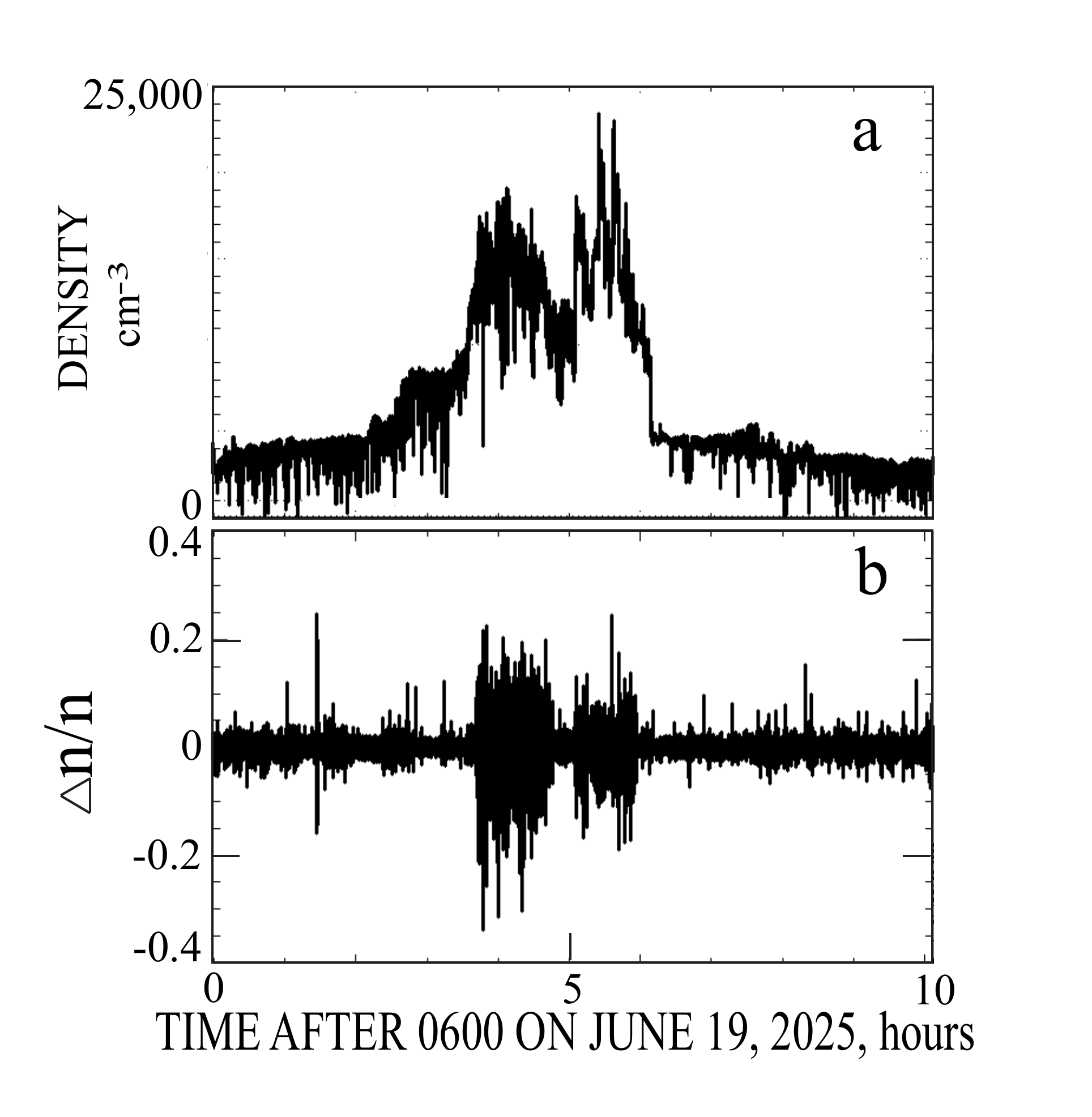}
   \caption{(5a) The plasma density; (5b) the density fluctuations $\Delta n/n$. That the density fluctuations were large shows that the plasma was turbulent. Both quantities were determined from fitting the spacecraft potential to the logarithm of the density determined from the quasi-thermal noise.}
    \label{fig:fig5}
\end{figure}

\section{Summary}
The Parker Solar Probe traversed a coronal pseudo-streamer at 10 solar radii. The identification of the structure as a pseudo-streamer is confirmed by   i) a double-peaked electron density structure, and (ii) invariant magnetic field polarity across the structure. In the plasma rest frame, electric fields as large as $400 \text{ mV/m}$ were observed. These fields were not associated with an $E \times B$ flow because such components of the field were removed when the prime frame was created. This prime electric field was supported by terms in the Generalized Ohm’s Law that have current densities as large as $1 \text{ mA/m}^2$ as an upper bound. The resistive and pressure gradient terms must have been non-negligible given that the normalized density fluctuations reached $0.3$ and the density curve had a non-zero slope during large electric field events.

%%%%%%%%%%%%%%%%%%%%%%%%%%%%%%%%%%%%%%%%%%%%%%%%%%%%%%%%%%%%%%
\begin{acknowledgements}
This work was supported by NASA contract NNN06AA01C. The authors acknowledge the extraordinary contributions of the Parker Solar Probe spacecraft engineering team at the Applied Physics Laboratory at Johns Hopkins University. The FIELDS experiment on the Parker Solar Probe was designed and developed under NASA contract NNN06AA01C. OVA was supported by NASA contracts 80NSSC22K0433, 80NNSC19K0848, 80NSSC21K1770, and NASA’s Living with a Star (LWS) program (contract 80NSSC20K0218).
\end{acknowledgements}
\bibliographystyle{aa}
\bibliography{references}
\newpage

%%%%%%%%%%%%%%%%%%%%%%%%%%%%%%%%%%%%%%%%%%%%%%%%%%%%%%%%%%%%%%%
% Appendices must be placed after   \end{thebibliography}
% They will be placed automatically on a new page.
%%%%%%%%%%%%%%%%%%%%%%%%%%%%%%%%%%%%%%%%%%%%%%%%%%%%%%%%%%%%%%%
\begin{appendix}
%%%%%%%%%%%%%%%%%%%%%%%%%%%%%%%%%%%%%%%%%%%%%%%%%%%%%%%%%%%%%%%
% In the PDF output, floats should be placed
% under their own appendix, not before the title, nor after the
% title of the next appendix.

% In short appendices, onecolumn floats (\figure*
% or \table*) will generate a blank page.
% To prevent this behaviour, a few examples are provided here. 

% In case you have a lot of floating objects for little text and the 
% LaTeX engine moves the floats away from their context, the command
% \FloatBarrier of the “placeins” package will empty the
% float buffer and place all stored floats in the continuity.

% If you still encounter problems with wide floats placement,
% just use the onecolumn environment throughout the appendices.
%%%%%%%%%%%%%%%%%%%%%%%%%%%%%%%%%%%%%%%%%%%%%%%%%%%%%%%%%%%%%%%

%____________________________________________________________
%       Wide floats at the start of an appendix: first method
%-------------------------------------------------------------
% To prevent a blank page after the start of an appendix:
% - Switch to one \onecolumn first
% - Declare the section title
% - Declare the onecolumn float with the parameter [ht!]
% - Revert to \twocolumn at the end of the section

\end{appendix}
\end{document}